**Identifying Diversity, Equity, Inclusion, and Accessibility (DEIA) Indicators for Transportation Systems using Social Media Data: The Case of New York City during Covid-19 Pandemic**

**Fariha Nazneen Rista**
Graduate Research Assistant
School of Civil Engineering and Environmental Science
University of Oklahoma
202 W. Boyd ST., Norman, OK 73019
Email: fariha.rista@ou.edu

**Khondhaker Al Momin**
Graduate Research Assistant
School of Civil Engineering and Environmental Science
University of Oklahoma
202 W. Boyd ST., Norman, OK 73019
Email: momin@ou.edu

**Arif Mohaimin Sadri, Ph.D.**
Assistant Professor
School of Civil Engineering & Environmental Science
University of Oklahoma
202 W. Boyd St., Norman, OK 73019
Email: sadri@ou.edu
**(*Corresponding Author*)**

## ABSTRACT

The adoption of transportation policies that prioritized highway expansion over public transportation has disproportionately impacted minorities and low-income people by restricting their access to social and economic opportunities and thus resulting in residential segregation. Policymakers, transportation researchers, planners, and practitioners have started acknowledging the need to build a diverse, equitable, inclusive, and accessible (DEIA) transportation system. Traditionally, this has been done through survey-based approaches that are time-consuming and expensive. While there is recent attention on leveraging social media data in transportation, the literature is inconclusive regarding the use of social media data as a viable alternative to traditional sources to identify the latent DEIA indicators based on public reactions and perspectives on social media. This study utilized large-scale Twitter data covering eight counties around the New York City (NYC) area during the initial phase of the Covid-19 lockdown to address this research gap. Natural language processing techniques were used to identify transportation-related major DEIA issues for residents living around NYC by analyzing their relevant tweet conversations. The study revealed that citizens, who had negative sentiments toward the DEIA of their local transportation system, broadly discussed racism, income, unemployment, gender, ride dependency, transportation modes, and dependent groups. Analyzing the socio-demographic information based on census tracts, the study also observed that areas with a higher percentage of low-income, female, Hispanic, and Latino populations share more concerns about transportation DEIA on Twitter.

## INTRODUCTION AND MOTIVATION

The transportation system is essential in providing people with a variety of options for getting to their desired destinations, and it has a significant impact on their quality of life *(1)*. With an ever-increasing influx of immigrants *(2)*, the United States of America is becoming more racially and ethnically diverse, necessitating a paradigm shift in transportation planning system to address diversity, equity, inclusion, and accessibility (DEIA) of its residents from all socioeconomic backgrounds. Traditional transportation planning lacks the concept of DEIA, thus making the system inaccessible to marginalized, underserved, and vulnerable communities.

An efficient transportation system should have three elements─ safety, mobility, and accessibility. However, most of the transportation agencies in the United States (U.S.) have safety and mobility as their primary target areas. Although transportation is the only means of accessing social and economic activities, accessibility has consistently been overlooked in traditional transportation decision-making processes, affecting individuals' ability to meet their needs and participate in civil society, which leads to social inclusion *(2)*. For example, the current car-based roadway infrastructure has been a barrier to accessibility for many people, particularly those who do not have regular access to a car and rely on public transportation for commuting. For these transit dependents, the continued availability of public mass transit is vital for access to jobs, schooling, medical care, and other necessities of life *(3)*. As a result, while the current transportation system meets mobility targets, it fails to address accessibility requirements from a social perspective, resulting in social imbalance and inequity.

Transportation planning that is inclusive, diverse, and equitable has the potential to ensure accessibility for all classes of people and change the way residents experience urban space in their daily lives. Therefore, it is imperative that organizations responsible for the transportation system planning, recognize the concept of DEIA and implement it at every level of their service and organization. The bright side of the picture is that the U.S. government has started to acknowledge the concept and has therefore enacted policies that require transportation agencies to implement DEIA in all aspects *(4)*. Therefore, rather than identifying the importance of DEIA, this study focuses on identifying the ways to measure inequity, exclusion, and inaccessibility in the transportation system of the U.S.

As the goal of the traditional transportation decision making process has been to achieve increased mobility *(5-7)*, mobility indicators, i.e., the technical and physical dimensions of transportation such as traffic speed, level of service, etc., are still predominant in transportation planning, completely ignoring the social dimension of it, i.e., diversity, equity, inclusion, and accessibility. Therefore, the indicator of DEIA is not established yet. Litman suggests that the quality of available transport options, average trip distances, and costs per trip could be indicators of accessibility and suggests a public survey to measure the indicators *(8)*. However, collecting survey responses is time-consuming and costly. Moreover, the opinion of survey respondents can change by the time the survey is complete. Social media platforms (SMPs) (i.e., Facebook, Twitter, Reddit, Instagram, etc.), on the other hand, can provide us with a faster way to acquire real-time data *(9)* and can be a potential metric for capturing the DEIA status of the transportation system.

According to the US Census Bureau, 84% of US households own a cell phone, and 78% own a desktop or laptop computer *(10)*. The Covid-19 outbreak saw more than 29.7% more users spending 1-2 hours per day on social media, with 20.5% using SMPs 30 minutes to an hour more *(11,12)*. This makes social media a viable data source for user-generated content. Twitter recently launched its Academic Application Programming Interface (API) *(13),* which provides a full history of public conversation *(14),* thus making Twitter a reliable social media data source.

In this study, Twitter data have been used to identify the latent indicators of DEIA in the transportation system that were not captured in traditional literature. The goal of this study is to assist the planners in measuring the DEIA performance of a community in a fast and efficient way. Therefore, the specific research questions (RQ) for our study are:

*RQ1: Can we identify latent DEIA indicators using public reactions and perspectives as expressed in social media?*
*RQ2: Can we identify the critical locations where the public reacts negatively about transportation DEIA issues?*

Identifying the sources of inequity arising due to the inaccessibility of transportation system in a society is a challenging problem as the sources may be interconnected. To explain this complex network, we have generated a conceptual framework (**Figure *1***). For simplification, we have broadly classified the communities into two categories─ rich communities and vulnerable communities. However, by "rich community", we identify the neighborhoods where people have access to various facilities and services such as schools, hospitals, groceries, restaurants, vacation trips, gymnasiums, playgrounds, banks, etc. Due to increased accessibility and better transportation connectivity, the housing costs in such areas are high. Sanchez et al. showed that Whites have a poverty rate of only 5 percent, compared to 22 percent for African Americans, 20 percent for Latinos, and 10 percent for Asian Americans *(15)*. Therefore, we assume that white people predominantly live in the rich areas of this framework. Households in such areas own at least one car, which is why people from such communities can fully and independently access the urban facilities.

On the other hand, the "vulnerable community" represents the group of people who cannot avail full access to the urban facilities compared to the "rich community". These groups have been ignored in the conventional planning and decision-making processes. Based on the related literatures till date, we have identified a few groups who fall in this category: (a) Low income and zero-vehicle households, (b) Dependent groups such as disabled people, people with age 65+ and age 15- , (c) Bicycle riders, (d) Pedestrian, (e) Minority groups such as Hispanic, Latino and Black or African American population, Women, LGBTQ+ population.

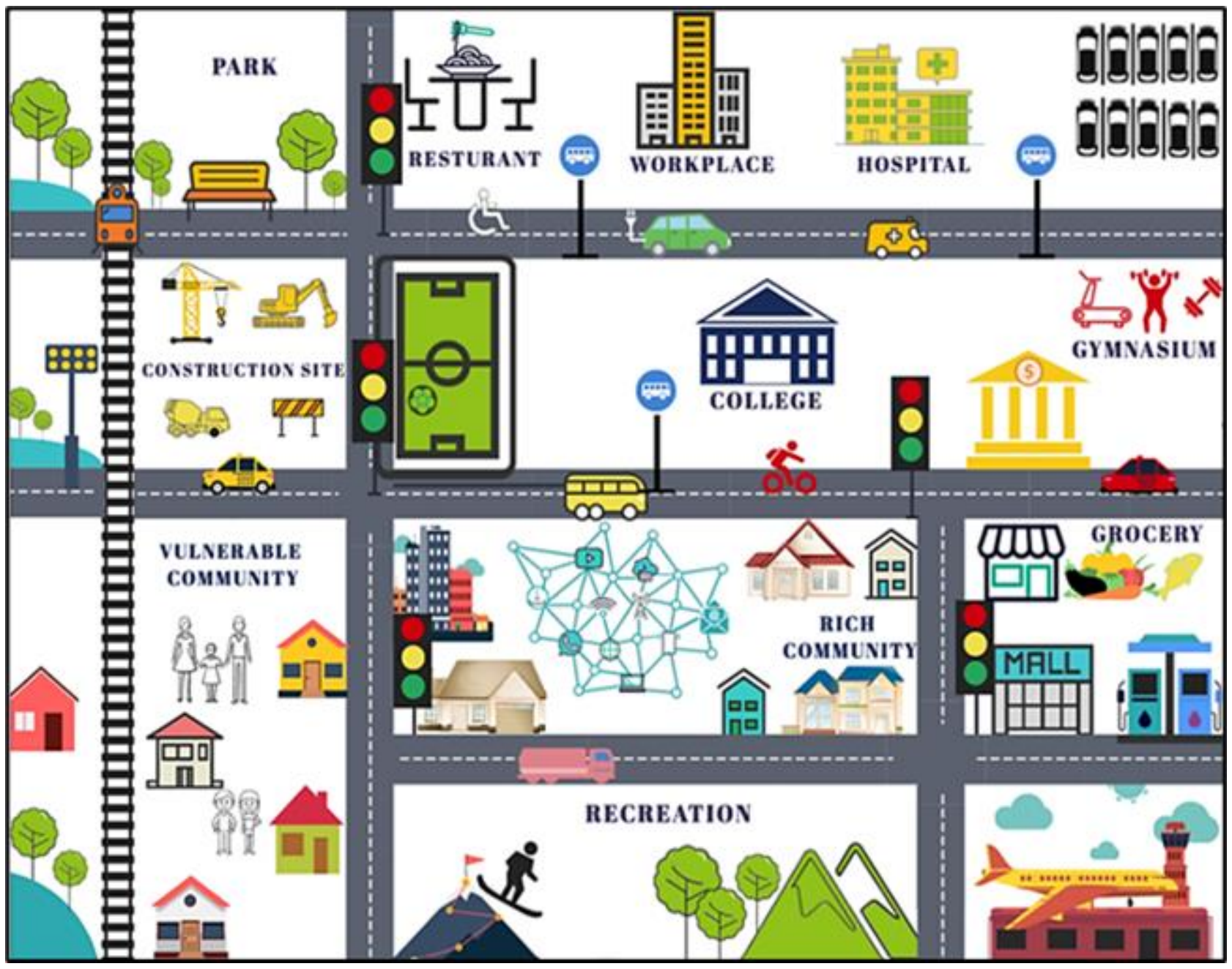

**Figure 1: Conceptual framework showing potential sources of DEIA issues in transportation system**

Literature shows that the largest percentage of public transportation users are minorities with low to moderate incomes *(15)*. In the U.S., transportation cost is the highest cost for households after housing costs, and it continues to rise. People who are displaced by rising property values frequently have few alternative housing options and may end up moving farther from their places of employment and social connections- a situation worsened by limited transportation options *(15)*. The disparity continues in the workplace as well. Women are paid lesser than their male counterparts. Therefore, the idea of inclusion needs to be incorporated from the organizational level. The highway-based transportation planning also fails to address the need for bicycle mode, which is an affordable alternative for individuals who cannot afford the cost of an automobile *(16)*. Almost 25 percent of Black households and 20 percent of Hispanic households lack a broadband internet connection causing a social inequity of information access. Based on these past works of literature, we have adopted several hypotheses (H) for our research:

*H1: Netizens discuss the DEIA issues they face on social media.*
*H2: Marginalized travelers (i.e., low-income females) suffer more from lack of inclusion.*
*H3: Locations with more diverse neighborhoods experience higher DEIA issues.*
*H4: Minority groups suffer more from inequity and exclusion*

## RELATED WORK

In recent years, social equity in transportation has been gaining more and more attention from transportation researchers, Transportation Research Board (TRB), National Academies, state Department of Transportation (DOT), U.S. DOT, American Society of Civil Engineers (ASCE), and the private sector to eliminate adverse impacts experienced by underrepresented and marginalized travelers and construction workers among others *(17-28)*. This is evident from Transportation Research Board (TRB) that recently identified complementary pathways to achieve DEI goals and set priorities at the local, tribal, state, regional, and federal levels *(29)*. However, earlier focus on diversity *(30,31)* has now shifted to equity considerations to reflect more on inequities due to ethnicity, language and income among others *(32)*.

Many researchers in recent times have been voicing about the need of transportation DEIA to eradicate the residential segregation. However, no one talked about the metric to measure it, except a few researchers. For example, Litman, in his research, has provided an overview of transportation equity concepts, methods to measure equity and incorporate it in planning analysis *(8)*. A research team at the Center for Urban Transportation Research (CUTR) at the University of South Florida, offered a two-fold solution to incorporate equity in traditional planning. They established an equity audit tool that identifies the community transportation needs from an equity perspective and a equity scorecard tool that prioritizes projects based on a score derived from the needs assessment result from equity audit tool *(33)*. However, such tools heavily depend on data that are obtained through public surveys.

Many studies recognized the potential and necessity of utilizing behaviorally enriched information on user activities and networks provided by emerging social media platforms. There is substantial empirical evidence that such large-scale data has the statistical power to reveal patterns of public opinion on ongoing societal issues *(34)*. For instance, many researchers used Twitter to study the service characteristics *(35,36)*, retweeting activity *(37,38)*, situational awareness *(39,40)*, online communication of emergency responders *(41,42)*, text classification and event detection *(43-47)*, devise sensor techniques for early awareness *(48)*, human mobility *(49,50)*, and disaster relief efforts *(51)*. Recently, transportation researchers used these data sources extensively for problems related to human mobility patterns *(52)*, origin-destination demand estimation *(53-57)*, activity-pattern modelling *(58-61)*, social influence in activity patterns *(62)*, travel survey methods *(63,64)*, transit service characteristics *(65)*, and crisis informatics *(66)* among others. However, there is not enough guidance how we can leverage such platforms to identify new or understudied DEIA factors related to transportation systems. As such, this is one of the first few studies in the literature that collects large-scale Twitter data obtained from locations around New York City (NYC) and analyzes transportation specific DEIA issues discussed by the residents around NYC on Twitter during the Covid-19 pandemic.

## DATA COLLECTION AND PREPARATION

We collected Twitter data from eight counties around NYC ─ Rockland, Westchester, Bronx, Queens, Kings, Richmond, New York, and Nassau. According to the 2020 census diversity index report, New York is one of the most diverse states in the United States, with a diversity index of more than 65 percent (**Figure 2**) with 19.5 percent of total population of Hispanic or Latin origin and 1.6 percent of total population of Black or African American race *(67)*. The research used New York City as a case study; however, this research can be scaled up or down based on necessity.

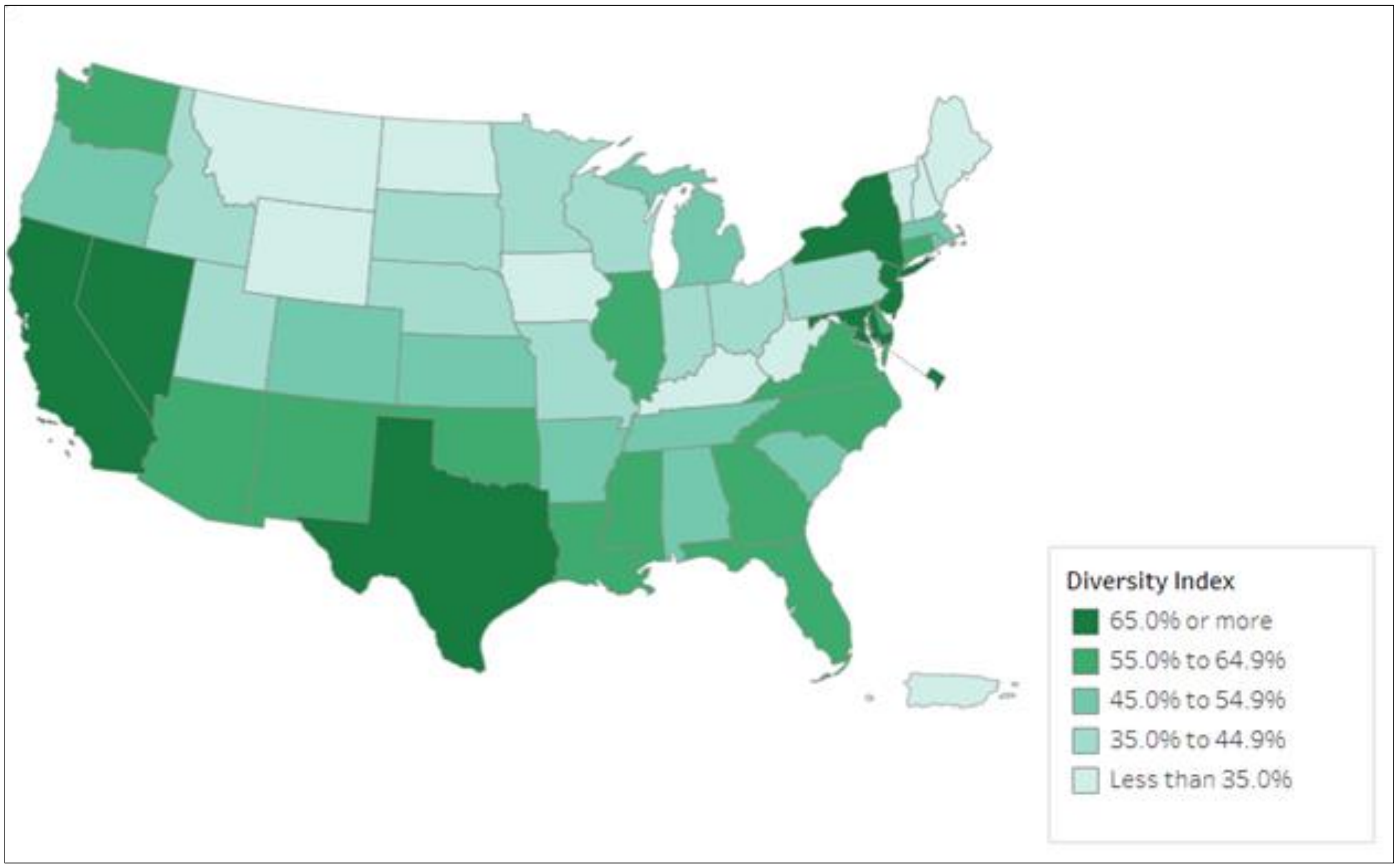

**Figure 2: Variation of Diversity in the USA**

### *Twitter Data*

Twitter data was collected using the recently launched Academic Application Programming Interface (API) *(13)*, which provides the full history of public conversation through a full-archive search endpoint *(14)*. Python programming language was used to collect the data, and associated Python libraries have been used. Geolocation-based search inquiry has been used to get data from the study area (**Figure *3***) from March 15 to May 15, 2020. The number of total collected tweets was 1.18 M tweeted by ~11.58k unique users.

Tweets retrieved from the academic track API contain additional information such as user id, username, profile information, and tweet location (coordinates of the neighborhood) along with the tweet text. Tweet texts, user information, and location information were considered for analysis in this study. Given the inherent ambiguity of tweets (e.g., non-standard spelling, inconsistent punctuation, and capitalization), additional preprocessing steps are performed to extract clean tweet text and username suitable for analysis. Noises (such as Html tags, character codes, emojis, and stop words) were removed from the text data and username, and tweets were tokenized, which is the process of breaking down an expression, sentence, paragraph, or even an entire text document into smaller units like individual words or phrases. Tokens are the names given to each of these smaller units.

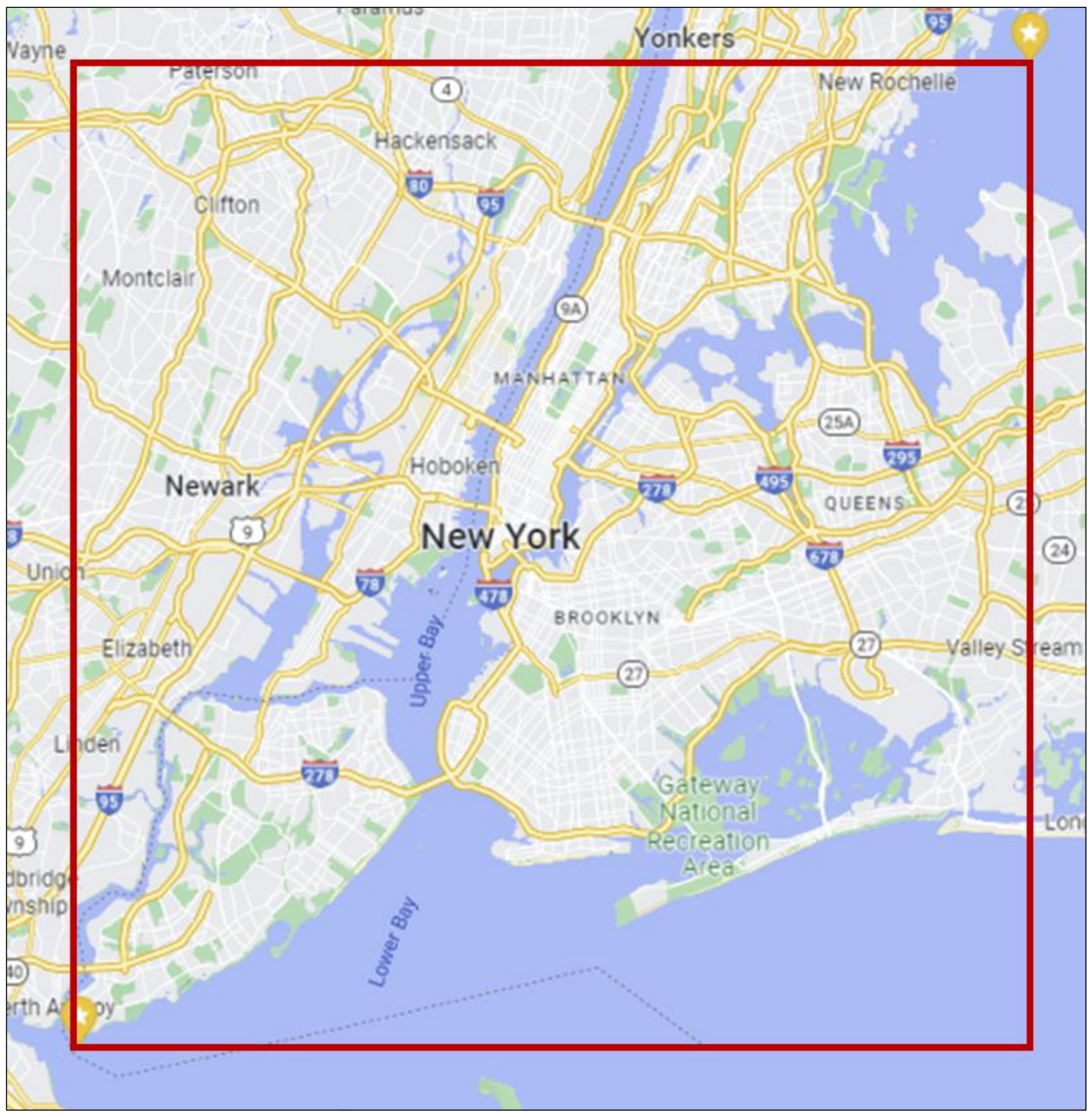

**Figure 3: Study area (inside the bounding box)**

Though the data was collected using geolocation-based search inquiry, it contained many geotagged tweets generated outside NYC. Tweets originating outside NYC, constrained by about (40.49, -74.25) and (42.14, -73.70) coordinates (**Figure *3***), were removed from the dataset. Moreover, Python's Pandas library is used to check for repetitions and ensure that all tweets in the data set are unique. The study considered tweets relevant to DEIA in transportation only. The relevance of a tweet was determined by identifying tokens within the tweet; more information on the steps and importance of relevance filtering can be found here *(48)*. To retrieve the tweets related to DEIA in transportation, two steps of relevance filtering were adopted using two keyword lists which are given below:

- **DEI related keywords (46 words):** *“diversity", "equity", "inclusion", "excluded", "inequity", "equality", "unequal", "inequality", "inequitable", "injustice", "unjust", "justice", "accessibility", "inaccessible", "affordability", "afford", "unaffordable", "affordable", "blacks", "racism", "race", "racial", "gender", "ethnicity", "discrimination", "discriminated", "wheelchair", "disabled", "disability", "ada", "hispanic", "black", "poor", "ethnic", "women", "gender", "deprived", "underprivileged", "disadvantaged", "underserved", "denied", "marginalized", "polarization", "aged", " income".*

- **Transportation related keywords (51 words):** *"flexwheels", "taxi", "scooter", "carshare", "carpool", "bike", "carclubs", "bicycle", "uberpool", "ebike", "multimodal", "passenger", "ride", "lyft", "uber", "bikeshare", "escooter", "mta", 'busstop', 'trains', "metro", "stations", "bus", "publictransport", "subwaystation", "train", "busstation", "autobus", "ridership", "transit", "subway", "publictransit", "station", "mtasubway", "car", "taxi", "drive", "traffic", "vehicle", "road", "automobile", "wheel", "bus", "subway", "transit", "amtrack", "rail", "flight" , "paralyzed", "roadtax"*

At first, DEI relevancy of the tweets was established if the tweet contained at least one of the DEI keywords identified for this study. After having the DEI relevant tweets, data was further filtered to keep the tweets concerning DEI in transportation using the transportation keywords list. Although this approach may filter out some relevant tweets, it ensures that all tweets involving these keywords were included in the filtered dataset for further analysis. At this step's end, a dataset of the relevant clean tweet was obtained with 19,194 tweets which is 1.62% of the original data set. **Figure 4** shows the distribution of data at different steps.

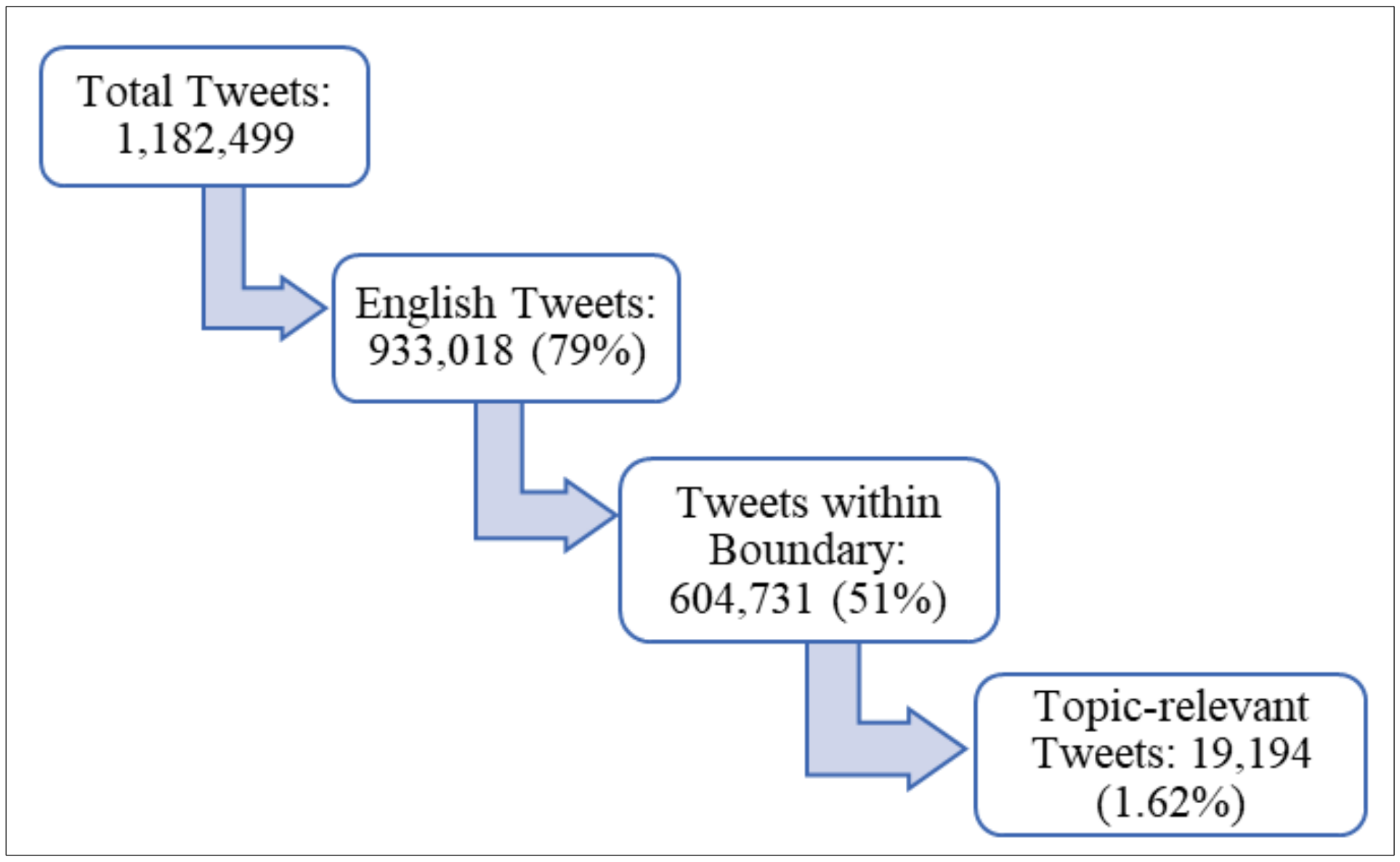

**Figure 4: Description of the dataset**

***Socio-demographic data***

The DEIA issues in the transportation domain vary across different communities based on the differences in socio-demographic properties. For instance, a community with a higher concentration of minority groups may experience greater inequity and accessibility issues than a neighborhood with a higher concentration of white Americans. To understand the correlation of socio-demographic features on DEIA in transportation, a few socio-demographic factors were collected at the census track level. At first, the census track information of the tweet originating places was generated by reverse geocoding using census geocode 0.5.2 python package *(68,69)*. The Twitter data was infused with demographic data collected for corresponding census tracts from the American community survey ACS 2020 database, which can be accessed via the US Census API *(70)*. The following four categories of socio-demographic information were considered for this study:

a) Per capita income from ACS
b) The proportion of the female population from ACS
c) The proportion of the Hispanic or Latino population from ACS
d) Proportion Black or African American population from ACS

## METHODOLOGY

We performed two forms of analysis on the clean relevant tweets (n= 19,194). At first, we performed text analysis where we applied unsupervised machine learning algorithms (bigram analysis, sentiment rating, topic mining) to identify transportation-related DEIA concerns in our study area. Secondly, we performed an investigative analysis using GIS mapping to explore the correlation between DEIA sensitivity and demographic factors. In brief, we wanted to explore how the topic and location of tweets vary depending on people's perception of DEIA. The key steps involved in the data analysis are summarized in **Figure *5***.

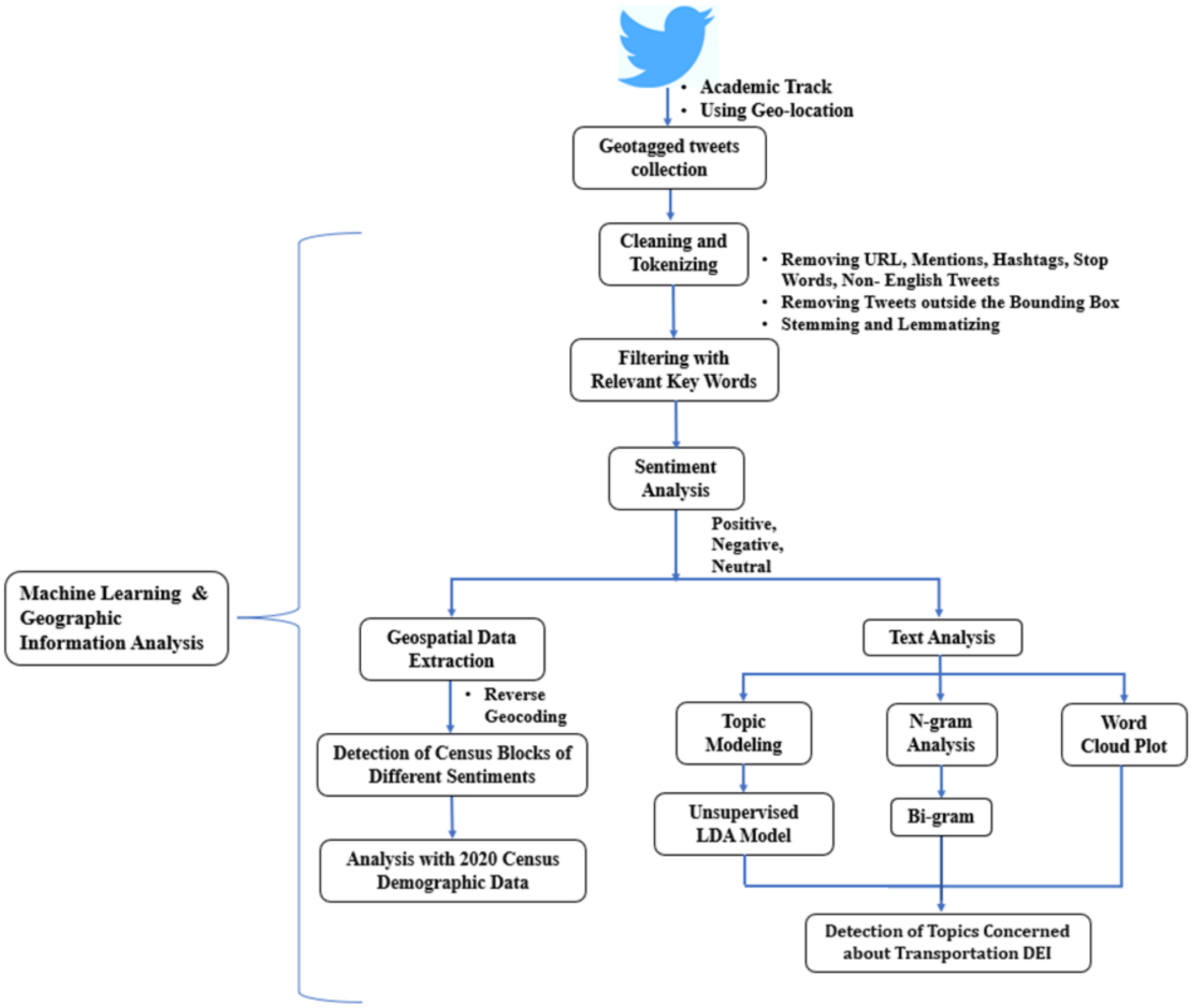

**Figure 5: Framework of the study**

### Word Bi-gram Analysis

Word bigram is a NLP concept that examines the likelihood of a pair of words appearing next in a series of words *(71)*. It can chunk pairs of words and predict the next forum word. The analysis separates the word sequence $w_1 \ldots w_N$ from the sequence of text observations $x_1 \ldots x_T$ for which the subsequent probability $P(w_1 \ldots w_N | x_1 \ldots x_T)$ reaches its maximum. The following can be written as Equation 1

$$arg\,max\,\{\mathrm{P}(w_1 \ldots w_N \mid x_1 \ldots x_T\,)\,.\ \mathrm{P}(x_1 \ldots x_T \mid w_1 \ldots w_N)\}, w_1 \ldots w_N \tag{1}$$

Where, $\mathrm{P}(x_1 \ldots x_T \mid w_1 \ldots w_N)$is the conditional probability of, given the word sequence $w_1 \ldots w_N$.. Using the conditional probabilities, we obtain the decomposition as Equation 2

$$\mathrm{P}(w_1 \ldots w_N\,) = \prod_{n=1}^{N} \mathrm{P}(\ w_n \mid w_1\ \ldots\ w_{n-1}\,) \tag{2}$$

We can divide a vocabulary of size W into a total of G word classes. The category mapping can be written as $G$: $w_0 \rightarrow G_w$. Each word w in the vocabulary is mapped to its word class $G_{w0}$. We use ($G_{v0}$, $G_{w0}$) to denote the equivalent class bigram for a word bigram (v,w). The equation for maximum likelihood estimate could be expressed as Equation 3

$$F_{bi}\ (ß) = \sum_{w} N(w)\ log\ \ N(w) + \sum\nolimits_{G_v,G_w} log\,\frac{N(G_v, G_w)}{N(G_v)N(G_w)} \tag{3}$$

Given, $F_{bi}$ = bigram maximum likelihood; $N$ = training corpus size; $W$= vocabulary size; *(u,v,w,x)* words in a text; ß = number of word classes *(71)*.

## Sentiment Analysis

The computational analysis of opinions, attitudes, and emotions is known as sentiment analysis or opinion mining. It attempts to deduce people's feelings from their linguistic expressions in a text. In this study, we used a Python package called VADER *(72)*, which detects the sentiment value of a short text, for analyzing the sentiments of relevant tweets about DEIA issues in transportation. Using a pre-defined list of words, VADER assigns a final compound score to each of the input words, which is the sum of all the lexicon ratings which have been standardized to range between -1 and 1 *(72)*. Following the methodology described in literature *(73)*, the tweets were classified into positive, neutral, and negative if the sentiment score ranged from -1 to 0, 0, and 0 to 1, respectively.

## Topic Mining

Topic modeling is used to segment, understand, and summarize a big dataset. Among many methods available, we particularly used the Latent Dirichlet Allocation (LDA) *(74)* model in our analysis. The purpose to choose this model is to discover the latent unknown topics related to transportation DEIA since a supervised model can only provide the known topics that it has been trained with. LDA is a generative probabilistic model where each document is assumed to contain a variable distribution of topics. In LDA, each document is assumed as a combination of topics, and each topic is assumed as a combination of words. To assign topics to each of the documents, LDA does the following actions:

i. *It starts with assuming K topics in the document and loops through m document and randomly assigns each word in the document to one of the K topics.*
ii. *For each document, it loops through each word $w_j$ and calculates:*
iii. *P (word $w_j$| topic $t_k$): Proportion of assignments to topic $t_k$ across all documents for a given word $w_j$*

*P (topic $t_k$| document $d_i$): Proportion of words in document di that are assigned to topic $t_k$*

iv. *Reassign topic 'T' to word $w_j$ with probability $p(t_k|d_i)$ * $p(w_j|t_k)$ considering all other words and their topic assignments*

The last step is repeated until we reach a steady condition where the topic assignments do not change any further. These topic allocations are then used to calculate the ratio of topics for each document. The procedure is shown in **Figure *6***.

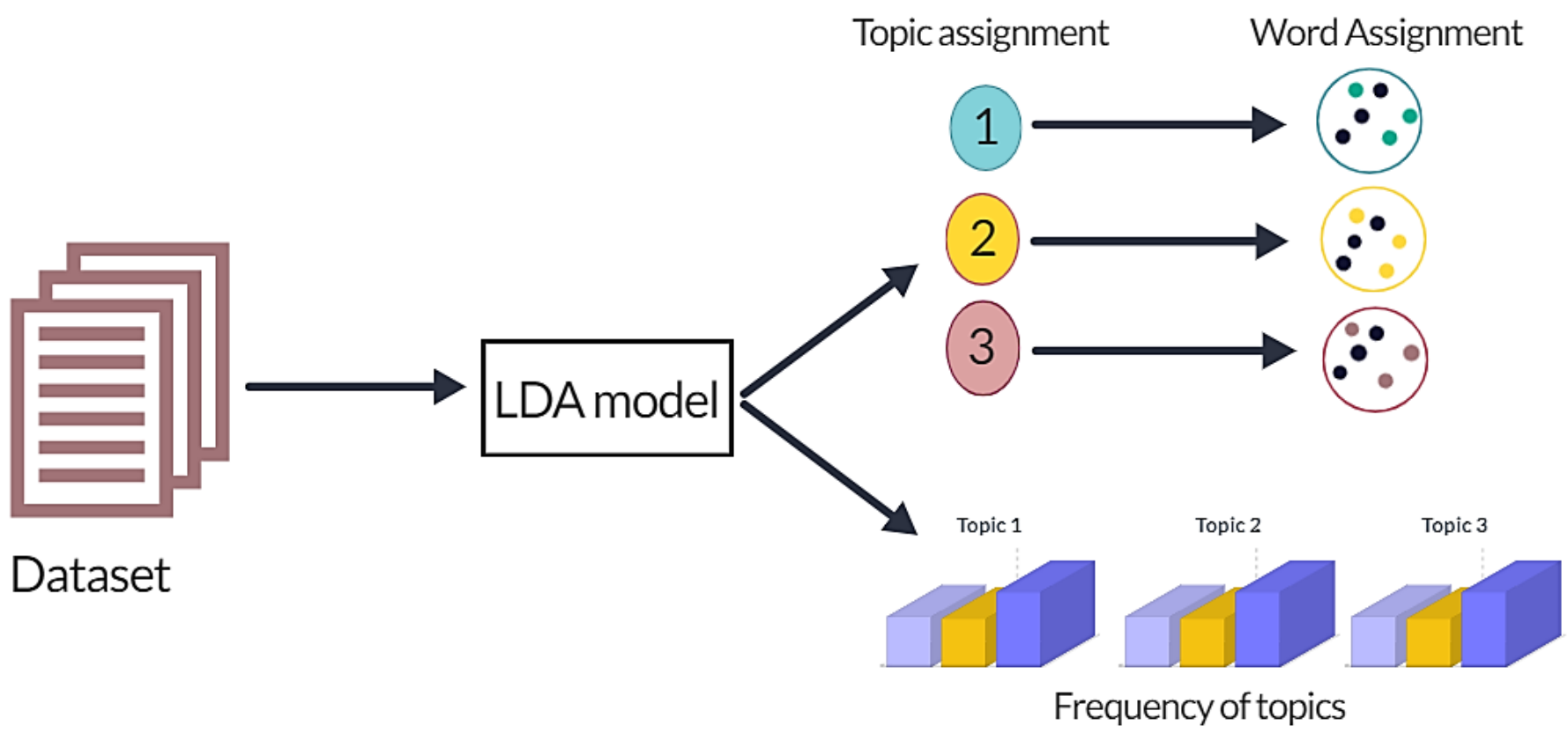

**Figure 6: Conceptual Figure of LDA model**

Therefore, topic modeling analysis was run for our study to investigate the hidden topics suggested by different sets of words present in the dataset. Rather than applying the model to the entire dataset, the dataset was separated first into positive, negative, and neutral segments based on the result of sentiment analysis, after which the model was applied separately to each segment. The approach can potentially identify the topics that citizens find constructive and are therefore positive about while also highlighting the topics that they find problematic and are thus negative about.

## Socio-demographic Correlation

To explore the correlation between DEIA sensitivity and demographic factors. The socio-demographic data and the Twitter data with sentiment information were mapped using ArcGIS *(75)*. The socio-demographic data were classified in the following manner:

**Table 1 Classification criteria for socio-demographic data.**

| | Low | Medium | High | Very high |
|---|---|---|---|---|
| Percentage of Female, F | F ≤ 30% | 31%≤F≤50% | F> 50% | - |
| Per Capita Income, I | I≤ $50k | $51k ≤ I ≤ $150k | I>$50k | - |
| Percentage of Hispanic/ Latino, HL | HL≤ 10% | 11%≤ HL≤ 25% | 21%< HL≤ 50% | HL> 50% |
| Percentage of African American, AF | AF ≤ 10% | 11% ≤ AF≤ 25% | 21%< AF≤50% | AF> 50% |

## RESULTS AND DISCUSSIONS

After data processing and cleaning, a total number of 19,194 relevant English tweets were obtained for further analysis. The tokens of words were utilized to determine the most frequently used words in our dataset. A word cloud was prepared using the top 50 frequently discussed tokens (**Figure 7)**. It is seen that the most frequent words are “black”, “race”, “station”, “train”, and “poor” which suggest the topic discussed among the netizens of New York during the early phase of lockdown, such as poverty, racism, transit dependency, etc.

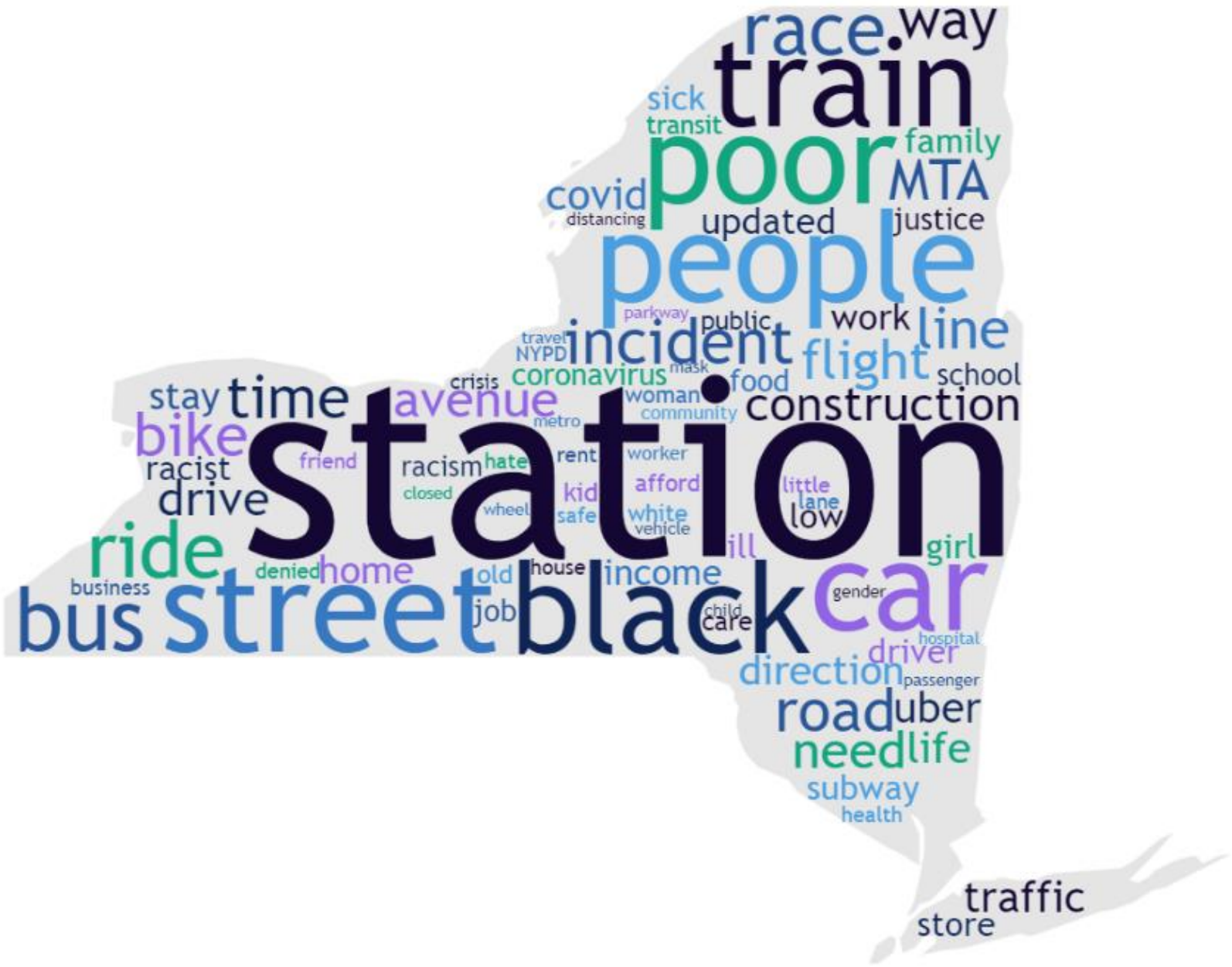

**Figure 7: Top 50 most frequent words in the dataset.**

Therefore, the text data can be a potential resource to identify topics of concern among the people of the study area, which are related to transportation DEIA. To fully utilized the data, two separate methods of topic mining were employed, as discussed in the following segments.

### Word Bi-gram Analysis

A word bigram is a pair of words that are next to each other in a text. **Figure 8** depicts which two words most frequently appeared together in the corpus. The study used a network to map the bigram words to gain a better understanding of their conversation in each community. The degree determines the color of each node in the network. The centered word has the highest degree, for example, black has a degree of 9 after white, which has a degree of 6, and these two words are accompanied by *people, person, brown, color, poor, American, movement, woman, men, community, folk, and man.* This indicates the presence of DEIA-related issues among people of color, females, and poor people. The analysis also showed the presence of protests regarding DEIA issues among the people. Most of the remaining words have one degree except for the *car, pay, rail, income, bike, blocked, and driver*. These words in the bigram are "buzzwords" that people use to talk about something and tell the story.

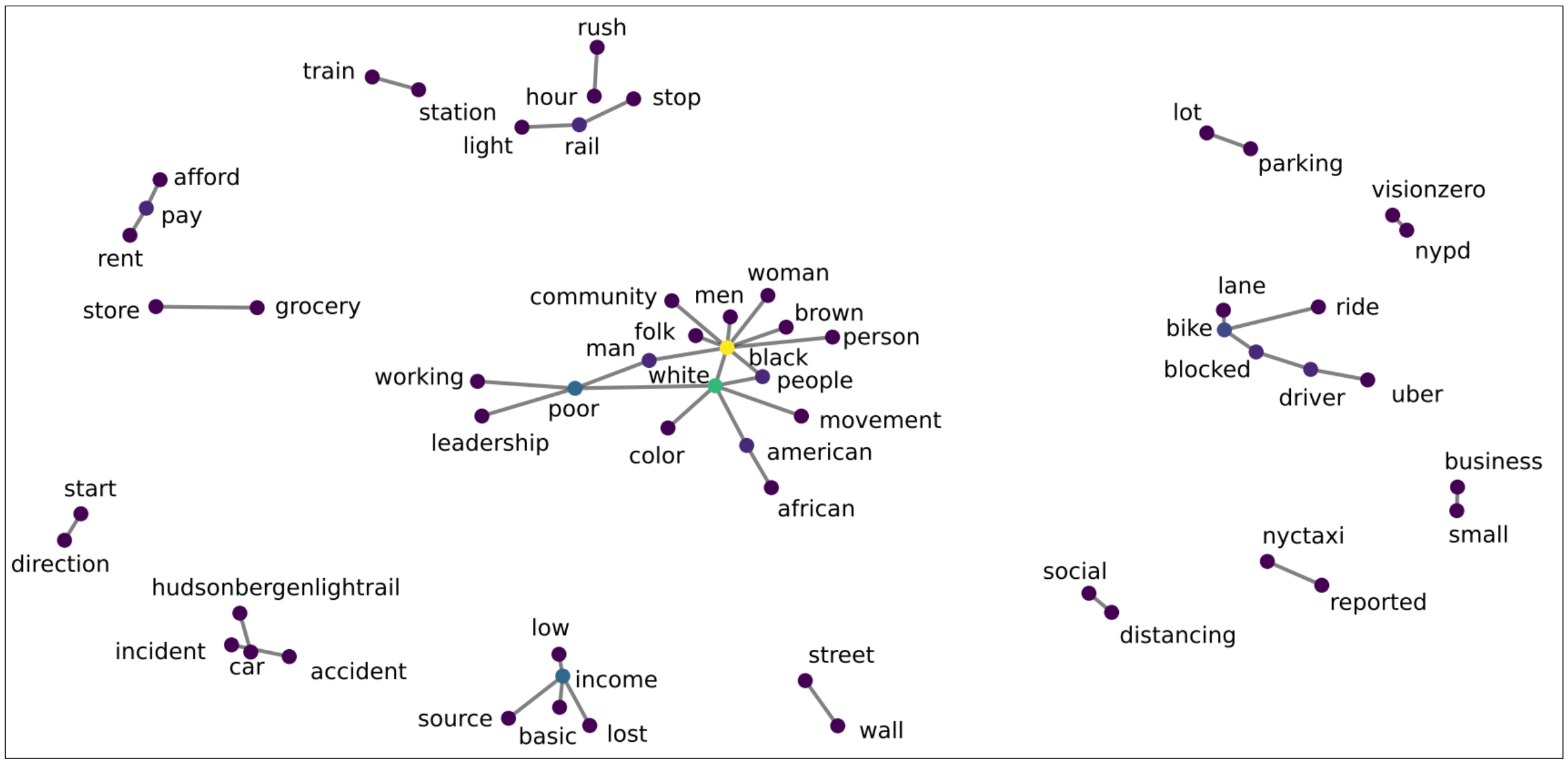

Figure 8: Results of Bigram Analysis

**Sentiment Ratings:**

**Figure 9 (a)** shows the output of the sentiment ratings, and **Figure 9 (b)** depicts the geolocation of tweets for various sentiment types. It was found that 33% of the tweets regarding the DEIA of transportation were positive, compared to 28% that were negative and 38% that were neutral. Being an aggregate analysis, the results did not provide enough insights regarding DEIA. That's why the topic of the following section was mined from the sentiments classified tweets.

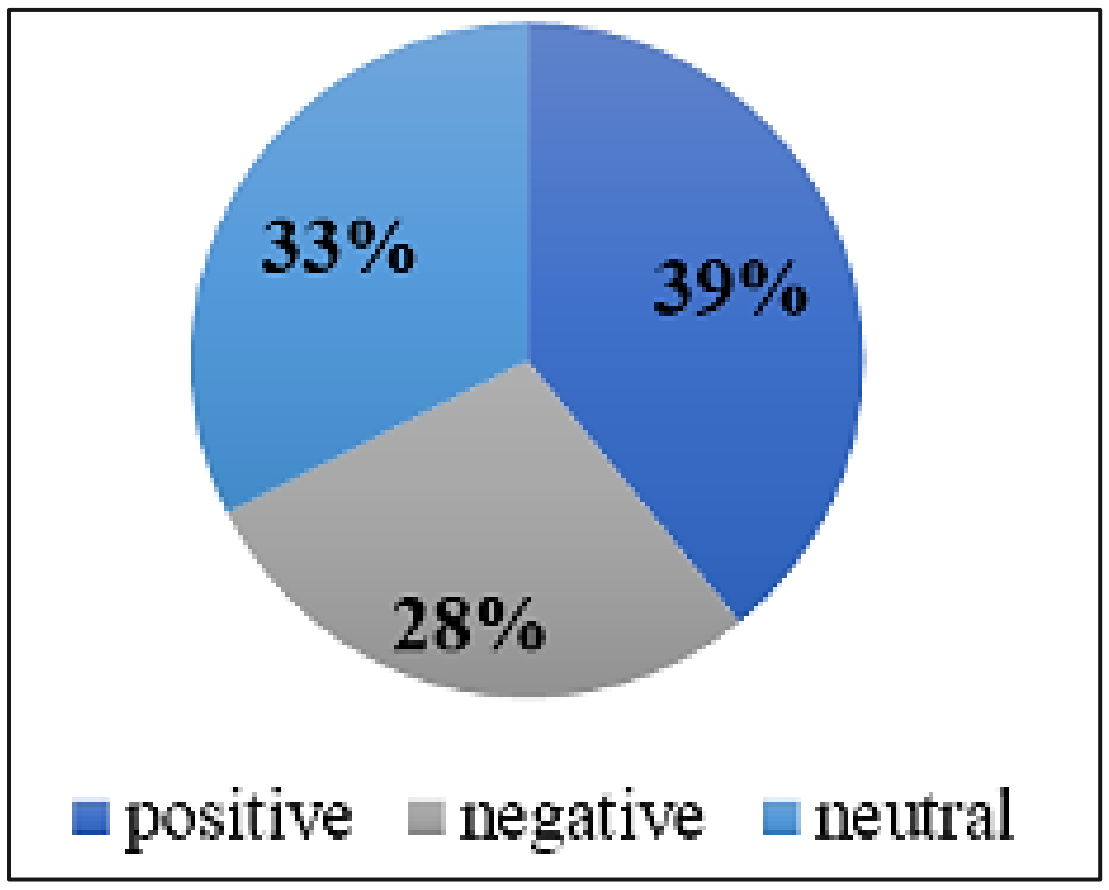

**Figure 9 (a): Result of sentiment ratings**

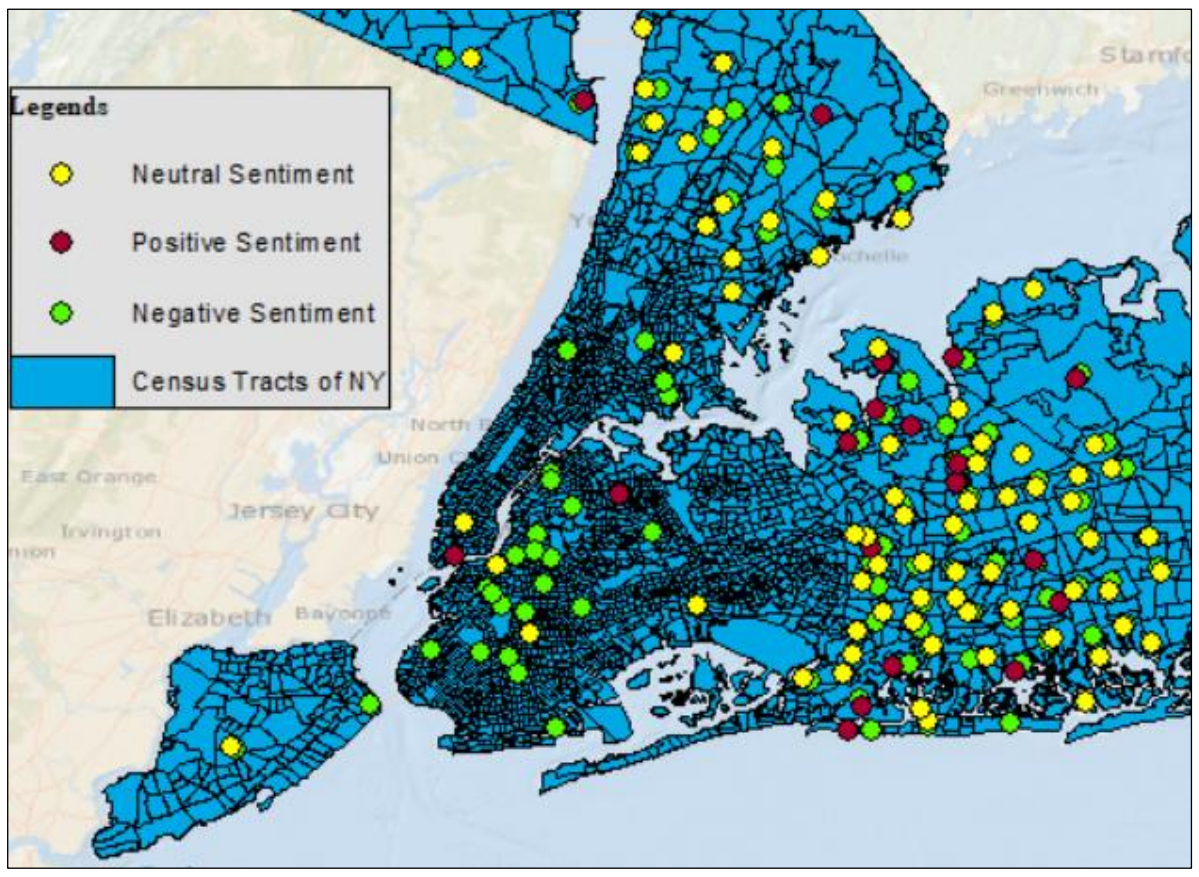

**Figure 9 (b): Spatial distribution of tweets**

**Topic Mining**

The optimum number of topics found for negative, neutral, and positive attitudes are respectively seven, five, and five. Among them, the topics which are relevant to transportation DEIA have been identified and presented in table 2. Table 2 shows the identified relevant topics and the relevant words contributing to those topics.

**Table 2 Transportation DEIA related most coherent topics.**

| **Sentiment Category** | **Topics (Topic Probability)** | **Most probable words in a topic** |
|---|---|---|
| Negative | Racism (0.528) | black (0.016), race (0.015), hispanic (0.013), racism (0.010), woman (0.009) |
| | Ride sharing (0.079) | driver (0.027), uber (0.023), time (0.020), dangerous (0.020), late (0.018), crime (0.017) |
| | Dependent group (0.107) | Wheelchair (0.016), child (0.014), young (0.013), ride (0.012), drive (0.011), |
| | Income and employment (0.055) | low (0.024), homeless (0.020), poor (0.017), income (0.016), rich (0.015), job (0.013) |
| | Gender (0.05) | gay (0.020), gender (0.014), women (0.012), language (0.010), deny (0.008) |
| | Transportation modes (0.044) | bus (0.081), transit (0.018), train (0.010), flight (0.036), car (0.013) |
| Positive | Diversity (0.534) | black (0.016), gender (0.014), woman (0.013), neighborhood (0.013), race (0.010), Hispanic (0.009) |
| | Public transit (0.189) | bus (0.081), station (0.015), transit (0.014), car (0.012), line (0.011), train (0.007) |
| | Work engagement (0.158) | working (0.012), place (0.013), drive (0.012), store (0.011), sell (0.009) |

Three original tweets from our dataset are given here to demonstrate the categories:

1. *"We are all doing our best to prevent the spread of #COVID19 by staying home. But not all of us have this luxury. Thank you to those out there still working. Our essential retail workers, postal workers, @FDNY, @nypd, Transit Workers, & public employees need our support."* ─ Tweeted by username: Fernandez4NY on March 20, 2020.

2. *"Hell no! Not for a job! My dude they are replacing you the same day. #TAG SUM1 THAT WORKS FOR TRANSIT!? #MTA #NYC #notme #noti #dayjob #fireworks #publictrasportation @ NYC"* ─ Tweeted by CFlyersTV on March 20th, 2020

3. *"Well it's essentially systematic racism that's rooted in putting communities of color in areas that negativity affect their health, restrict their resources, or put them in vulnerable situations. It's rooted in land use policies dating back to the 1900s"* ─ Tweeted by username: NiicoleHill on May 8, 2020

The first tweet is an example of a positive tweet (sentiment score: 0.892) that talks about "work engagement". On the other hand, the second tweet is an example of negative sentiment (sentiment score: -0.498), which falls in the topic category, "Employment". Both the tweets were posted on the same day, March 20. The third tweet is an example of a negative tweet (sentiment score: -0.8689) which falls in the topic category "Racism". From table 2, it can be seen that people primarily discussed racism and gender negatively, which indicates their impact on transportation DEIA. Ridesharing and Transportation modes appeared as prominent topics among negative tweets indicating the presence of DEIA issues in these areas. People also talked about the DEIA concerns in transportation for the dependent group of people. Moreover, income and employment play an important role in transportation DEIA, which is indicated by the topic named income and employment. These results can be a potential source for planners to identify problems in a certain area. On the other hand, people were found to be speaking positively regarding the diversity in public transit and workplaces.

## Socio-demographic Correlation

In total, 119 census tracts were identified from where all the 19,194 tweets related to DEIA. Among them, negative sentiment tweets were generated from 99 census tracts, while positive and neutral sentiment tweets were generated from 76 and 74 counties, respectively. This tells the fact that people from some census tracts post both positive and negative tweets related to DEIA. These census tracts can be categorized as "highly sensitive" tracts for this study. To understand the demographic scenario of these locations, the location data are further overlaid on census demographic data at the census tract level to determine and categorize the demographic scenario of the locations. The tweet locations were laid over the colored maps of each type of demographic data to visualize the demographic scenario of the census tracts from where people were tweeting about DEIA within our study area. The results have been shown in the following **Figure 10.**

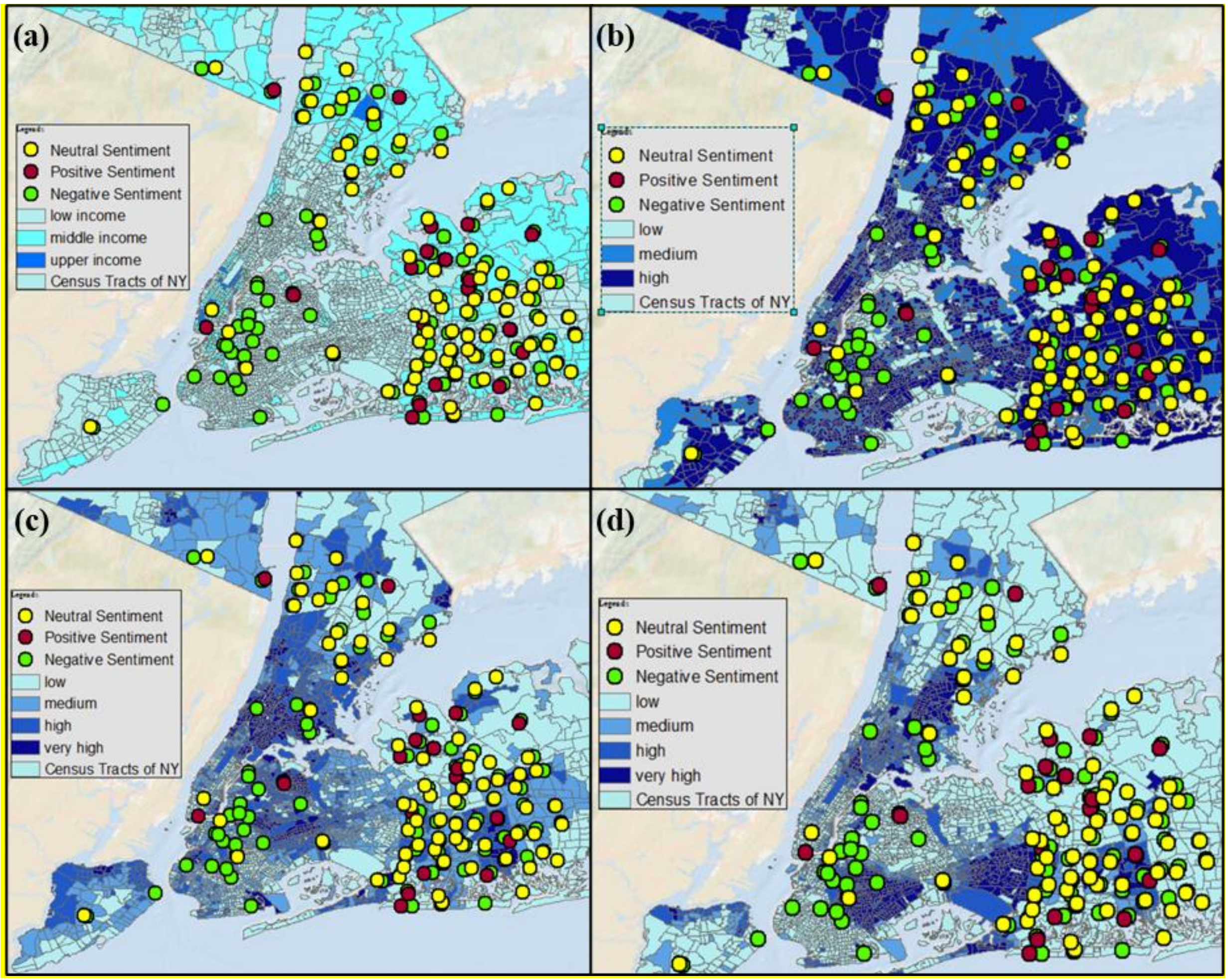

**Figure 10:** Spatial distribution of tweeting activity across different socio-economic groups. (a) Per capita income, (b) Female population, (c) Hispanic/ Latino population, (d) Black/ African American population

**Figure 11** shows the distribution of the tweets (percentage) according to different categories (e.g., low, medium) of sociodemographic factors (e.g., income, female proportion). The "Y" axis represents the percentage. It is found that 53% of tweets are coming from low-income group areas, and 45% of tweets are generated from medium-income group areas. However, only one percent of tweets come from high-income group areas. This indicates that people living in low-income and medium-income group areas talked more about transportation DEIA than people living in high-income group areas. This clearly shows the presence of transportation DEIA-related issues in low- and medium-income group areas. It supports our hypothesis H2.

In the case of females, most of the transportation DEIA tweets (74%) came from high female density areas, and the rest of the tweets came from medium female density areas. However, no tweets were generated from areas with low female density. This represents that the transportation DEIA-related issues appeared more in those areas where more female lives. This also supports our hypothesis H2. In the Hispanic or Latino population category, it is observed that 70% of the census tracts have a medium to a very high density of Hispanic or Latino population. Only 30% tweet came from areas where there is a low density of Hispanic or Latino population. This again supports our hypothesis H3.

However, in the Black or African American category, it is found that ~70% of tweets came from areas with a low density of Black population. This suggests that the Black population is more sensitive about DEIA issues in areas where they are minorities. This supports our hypothesis H4.

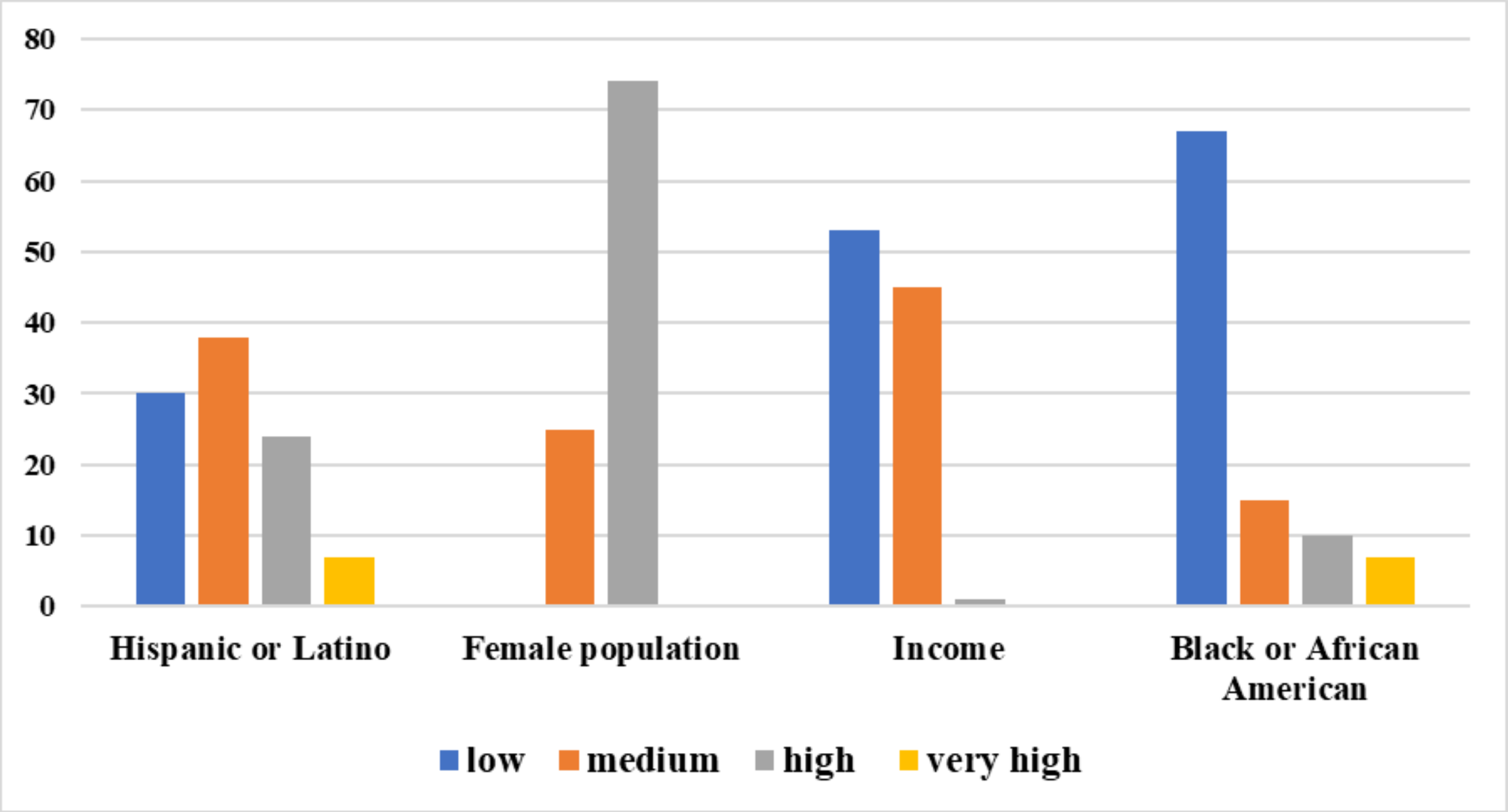

**Figure 11: Socio-demographic classification (by percentage) based on ethnicity, gender, income, and race in tweeting locations**

## CONCLUSIONS

Recently, transportation researchers have been leveraging social media data for issues including traffic incident and disaster management, activity pattern modeling, transit service assessment, and travel demand forecasting. However, the empirical literature is inconclusive regarding the use of social media data as a viable alternative to traditional sources to identify the latent diversity, equity, inclusion, and accessibility (DEIA) based on public reactions and perspectives as emanated through social media. This study demonstrated that geotagged Twitter data can potentially capture new insights about transportation equity, sense of inclusions, and the accessibility of diverse neighborhoods using the following methodologies—

- Negative sentiment resulting from VADER lexicon sentiment analysis can be a potential indicator of DEIA. Analyzing the negative sentiment tweets, we could detect the main concerns of people regarding DEIA within the study area.
- Applying reverse geocoding, we could detect the census tracts from where relevant tweets were generated and how many tweets are generated from which tract. Thus, we can find out the areas with the highest tweets about DEIA and mark the target areas accordingly. We also used the census tract data to explore the demographic information of these areas.

Following the scope and the specific research questions we raised in this study; we have the following conclusions from the study area:

- People tweeting about DEIA were more negative about issues like racism, income, unemployment, gender, ride dependency, transportation modes and dependent group. It can be concluded that the area suffered from these issues most acutely.
- Locations where residents were more sensitive about DEIA, had a high percentage of poor income people.
- People are more sensitive about inequity of transportation system where percentage of female population was high.
- People from locations with higher Latino and Hispanic population are more sensitive about DEIA of transportation system.
- Minority groups suffer more from inaccessibility and inequity.

Twitter data is free and provides real-time information. It is one of the most popular microblogging sites for social media users and has an accessible API. Survey data is largely dependent on the responsiveness and emotional state of the public. Therefore, it can be biased in addition to being time-consuming and expensive. Although we only used Twitter data in this study, in the future, when data from other social media platforms are accessible, a cross-platform analysis could produce more comprehensive results.

However, Twitter data has several limitations. It is important to emphasize how representative tweets are. Although we used Twitter data to determine how individuals feel about diversity, equity, inclusion, and accessibility of the transportation system, given that not all citizens use Twitter, the data cannot accurately represent the general populace. Additionally, the number of tweets including coordinates decreased because of the 2015 change in policy on exact location sharing, thus decreasing the representativeness of Twitter data furthermore. However, Goodchild and Glennon showed in their research that despite these drawbacks, employing geotagged social media datasets had more advantages than drawbacks *(76)*. Twitter data still have advantages over traditional survey data in terms of volume and pace especially through their recently released academic track that releases more comprehensive data than ever before.

The application of our research is significant. It can help transportation planners, designers, and practitioners identify potential sources and locations of diversity, equity, inclusion, and accessibility issues specific to the transportation infrastructure system within a given area and take necessary steps.

## ACKNOWLEDGMENTS

This material is based upon work supported by the National Science Foundation under Grant No. IIS-2027360. However, the authors are solely responsible for the findings presented in this study. The analysis and results section is based on the limited Twitter dataset and authors' opinions which cannot be expanded to other datasets without detailed implementation of proposed methods. Other agencies and entities should explore these findings based on their application/objectives before using these findings for any decision-making purpose. Any opinions, findings, conclusions, or recommendations expressed in this material are those of the author(s) and do not necessarily reflect the views of the National Science Foundation.

## AUTHOR CONTRIBUTIONS

Fariha Nazneen Rista: Conceptualization, Methodology, Formal analysis, Visualization, and Draft preparation. Khondhaker Al Momin: Data collection, Methodology, Review & editing. Arif Mohaimin Sadri: Conceptualization, Data collection, Supervision, and Review & editing. All authors reviewed the results and approved the final version of the manuscript.